\documentstyle[12pt]{article}
\def\ba{\begin{eqnarray}}
\def\ea{\end{eqnarray}}

\def\lb{\label}
\def\be{\begin{equation}}
\def\ee{\end{equation}}

\begin{document}

\begin{center}
\Large{ GENERALIZED GRASSMANN ALGEBRAS AND ITS \\
CONNECTION TO THE EXTENDED SUPERSYMMETRIC MODELS}
\end{center}

\vspace{.5cm}

\begin{center}
\large{A.P. ISAEV$^{*+}$, Z. POPOWICZ$^{**\diamond}$ and
O. SANTILLAN$^*$}
\end{center}
\begin{center}
${}^*$ Bogoliubov Laboratory of Theoretical Physics, JINR, \\
141 980 Dubna, Moscow Region, Russia \\
${}^{**}$Institute of Theoretical Physics, University of Wroc{\l}aw\\
pl. Maxa Borna 9, 50--204 Wroc{\l}aw, Poland
\end{center}

\vspace{1cm}

\begin{center}
ABSTRACT
\end{center}
It is shown that the fermionic Heisenberg-Weyl algebra with $2N=D$
fermionic generators is equivalent to the
 generalized Grassmann algebra with two fractional generators. The 2,3 and 4
dimensional Heisenberg - Weyl
 algebra is explicitly given in terms of the fractional generators. These
algebras are used for the
 formulation of the $N=2,3,4$ extended supersymmetry. As an example we
reformulate
 the Lax approach of the supersymmetric Korteweg - de Vries equation in terms of
the generators of the generalized
 Grassmann algebra.

\vspace{3cm}

${}^+$ E-mail: isaevap@thsun1.jinr.ru

\vspace{.3cm}

${}^\diamond$ E-mail: ziemek@ift.uni.wroc.pl

\newpage

\section{INTRODUCTION}
 Generalized Grassmann algebras have been introduced firstly in
 \cite{ABL} in the framework of
the 2D conformal field theories. Later these algebras were
rediscovered in the contexts of quantum groups \cite{FIK} and
fractional generalizations of supersymmetric quantum mechanics
\cite{Dur0}. Next, these algebras were used for the generalization
of the supersymmetry to the fractional supersymmetry \cite{FIK2},
\cite{AzM}, \cite{AD}, \cite{Dur}.

Interestingly the generalized Grassmann algebras (or their intimate analogs)
 have found applications
in  the construction of the finite dimensional (cyclic)
representations of the quantum groups, covariant quantum algebras
and zero-mode algebras in WZNW models (see e.g. \cite{FIK}, \cite{Isx},
\cite{Is}, \cite{AD}, \cite{DVT}).
The generalized Grassmann algebras should be also useful
for the construction of finite dimensional representations of
the quadratic diffusion algebras \cite{Ritt} which are employed in
the theory of one-dimensional stochastic processes with exclusion.
The fractional extensions of
the Virasoro algebra \cite{FIK2} utilizes generalized Grassmann algebras as well.
Recently generalized Grassmann algebras appeared  in the investigations of the
particle systems on linear lattices with periodic boundary conditions
\cite{Dobr} (see also \cite{Is}). In this paper we find the irreducible representations
of the Clifford algebras by realizing them in terms of the generalized Grassmann algebras. 

The main idea of the introduction of the generalized Grassmann
algebras is to replace the usual fermionic condition $\theta^2=0$
by the more general nilpotence condition $\theta^{p+1}=0$. The $p$
number  in the physical literature is usually  called the order of
parastatistics. Due to the big interest to such generalization it
is tempting to try to use these generalized algebras to the
supersymmetric solitonic equations.

In this paper we propose the  method of the  reformulations,  in
terms of the generalized Grassmann variables,  the usual extended
supersymmetric models. We concentrate our attention to the
supersymmetric Korteweg - de Vries (SKdV) equation only. However
our approach could be used to any $N=1,2,3,4$ supersymmetric
model as well.

We explicitly show that the fermionic Heisenberg-Weyl algebra with
$2N = D$ fermionic generators is equivalent to the generalized
Grassmann algebra with two fractional generators $\theta,
\partial$ where the order of parastatistics $p$ is $2^N-1$. We
explicitly construct the $2,3,4$- dimensional Heisenberg-Weyl
algebras in terms of the generalized Grassmann algebras with $p+1
= 4,8,16$. It appears that this Heisenberg-Weyl algebra is closely
related to the Clifford algebras in Euclidean spaces of the
dimensions $D=4,6,8$. One can use this observation to formulate
the $N=2,3,4$ extended supersymmetry. All of this allowed us to
formulate the $N=2$ SKdV equations in terms of the generalized
Grassmann algebras for $p+1=4$.

The paper is organized as follows: In Section 2 we recall the
basic facts about generalized Grassmann algebras. In Section 3 the
Weyl construction for the Heisenberg-Weyl and Clifford algebras is
discussed. In Section 4 we construct fermionic $N=2,3,4$
Heisenberg-Weyl algebras in terms of the generators of the $Z_4$,
$Z_8$ and $Z_{16}$ generalized Grassmann algebras. In Section 5 we
discuss the example of the $N=1,2$ SKdV equations rewritten in
terms of $Z_2$ and $ Z_4$ graded algebra respectively. The Section
6 contains conclusions.

\section{GENERALIZED GRASSMANN ALGEBRA}

 Consider a deformed oscillator algebra with
two generators $\theta$ and $\partial$ satisfying\cite{ABL},
\cite{FIK}, \cite{AzM}
  \be
  \lb{qdef1}
 \partial \, \theta - q \, \theta \, \partial = I \; ,
  \ee
  \be
  \lb{qdef}
  \theta^{p+1}=\partial^{p+1}=0 \;\; \Rightarrow \;\; q^{p+1}=1 \; ,
  \ee
 where $p$ is an integer. Note that the complete basis of this algebra
is given by the $(p+1)^2$ elements $(\theta^n \, \partial^m)$ $(0
\leq n,m \leq p)$ and this algebra is isomorphic to the algebra of
matrices $Mat(p+1)$ \cite{FIK}, \cite{FIK1}. The grading in this
algebra is
  \be
  \lb{grad}
  deg \, (\theta^n \, \partial^m) = n-m \; .
  \ee
 It is convenient to introduce the grading operator \cite{FIK}
  \be  \lb{gg}
 \omega = \partial \, \theta - \theta \, \partial \;\; \Rightarrow \;\;
 \omega^{p+1} = 1 \; ,
  \ee
 which defines the automorphisms
  \be
  \lb{ggg}
 \omega \, \theta \, \omega^{-1} = q \, \theta  \; , \;\;\;
 \omega \, \partial \, \omega^{-1}  = q^{-1} \, \partial \; .
  \ee
 Using (\ref{qdef1}) and (\ref{gg}) we obtain useful relations
  \be
  \lb{usr}
  \theta \, \partial = \frac{\omega-1}{q-1} \; , \;\;\;
  \partial \, \theta = \frac{q \, \omega-1}{q-1} \; .
  \ee

 The algebra only defined  by (\ref{qdef1}), but not  (\ref{qdef}),
has been firstly considered in \cite{ArCo}. This algebra is
closely connected to the algebra of Macfarlane - Biedenharn
quantum oscillators \cite{MaBi}. For the nontrivial case,  when
$q$ is root of unity, this connection has been discussed in
\cite{Is}.

 A characteristic equation $\omega^{p+1} = 1$ for the grading
 operator (\ref{gg}) can be rewritten in the form
  \be
  \lb{ggo}
 (\omega - 1) \, (\omega - q) \dots (\omega - q^p) =
 \prod_{n=0}^{p} \, (\omega - q^n) = 0 \; ,
  \ee
 and one can introduce the projection operators
  \be
  \lb{proj}
 P_k = \prod_{n=0, n \neq k}^{p} \, \frac{(\omega - q^n)}{(q^k -
 q^n)} \; , \;\;\; (k = 0, \dots, p) \; .
  \ee
 These projectors satisfy the following identities
  \be
  \lb{proj1}
  \begin{array}{c}
    \sum_{k=0}^{p} \, P_k = 1 \; , \;\;\; P_n \, P_k = \delta_{nk} \, P_k \; ,   \;\;\;
   \omega \, P_k =P_k \, \omega = q^k \, P_k,    \\
  P_0 \, \theta = 0 \; , \;\;\; \theta \, P_{k-1} = P_k \, \theta \; , \;\;\;
      \theta \, P_p = 0, \\
 \partial \, P_0  = 0 \; , \;\;\; P_{k-1} \, \partial = \partial \, P_{k} \; , \;\;\;
    P_p \, \partial = 0.  \\
  \end{array}
   \ee
 In view of the nilpotence conditions (\ref{qdef}) one can deduce
  \be
  \lb{nilp}
  \begin{array}{l}
 \theta^{p+1-n} \, \partial^{p+1-n} \, \partial^n = 0 \;
 \Rightarrow 
  \;\; \prod_{k=0}^{p-n} \, (\omega - q^k) \,
 \partial^n = 0.  \\ \\
 \theta^n \, \theta^{p+1-n} \, \partial^{p+1-n} = 0 \;
 \Rightarrow 
 \;\; \theta^n \, \prod_{k=0}^{p-n} \, (\omega - q^k) = 0.
  \end{array}
  \ee
 It  means that
  \be
  \lb{nilp1}
 \theta^n \, P_m = 0 = P_m \, \partial^n \; \;\;\; ( p-n < m \leq
 p). \;
  \ee

   A generic function $u$, with $\theta$ as an argument, can be expanded
 in power series as
  \be
  \lb{pow}
  u(\theta)= u_{0} + \theta \, u_{1} + \dots + \theta^{p} \, u_{p}.
  \ee
 where $u_i$ are in general noncommutative coefficients.
 The left action of $\theta$ on $u(\theta)$ is given by
  \be
  \lb{act}
  \theta \, u(\theta) = \theta \, u_{0} + \theta^{2} \, u_{1} + \dots +
  \theta^{p} \, u_{p-1}.
  \ee
 To define the left action of the operator $\partial$ on the
 function $u(\theta)$ (\ref{pow}) we find, using the relation
 (\ref{qdef1}), that
    \be
  \lb{der}
 \partial \, \theta^{n}
 =(1+q+q^{2}+...+q^{n-1}) \, \theta^{n-1}+ q^{n} \, \theta^{n} \, \partial =
 (n)_{q} \, \theta^{n-1} + q^{n} \, \theta^{n} \, \partial,
 \ee
 where the q-number $(n)_q$ is
  \be
  \lb{qnum}
 (n)_{q}=\frac{(1-q^{n})}{1-q}.
  \ee

 It is clear that $\partial$ can be considered as a generalization of the
 ordinary derivative and  using (\ref{der}) we obtain
  \be
  \lb{der2}
 \partial \,( u(\theta) ) = u_{1}+ (2)_{q} \, \theta \, u_2 +
 \dots + (p)_{q} \, \theta^{p-1} \, u_{p}.
  \ee
 The equation $q^{p+1} = 1$ is a consequence
of the nilpotence condition $\theta^{p+1}=0$ (\ref{qdef}) and
relation (\ref{der}) for $n=p+1$. If in addition we require that
$\partial (\theta^n) \neq 0$ for $n < p +1$, then $q$ should be a
primitive root of unity ($q^n \neq 1$ for all $n < p +1$) and one
can choose $q= \exp(2 \pi i/(p+1))$. In this case, generators
$\theta$ and $\partial$ (\ref{qdef1}), (\ref{qdef}) are reduced to
the ordinary fermionic ($p=1$) and bosonic ($p \rightarrow
\infty$) creation and annihilation operators.

 Equations (\ref{act}) and (\ref{der2}) give a matrix
 representation for $\theta$ and $\partial$. Indeed,
 arbitrary function $u(\theta)$ (\ref{pow})
 can be realized as $(p+1)$ dimensional vector
  \be
  \lb{mrepr}
 u = \left(\begin{array}{c}
   u_{0} \\
   \vdots \\
   u_{p}
 \end{array}\right) \; , \;\;\;
  u_k = < k | u >.
  \ee
 To reproduce the actions
of $\theta$ and $\partial$ on the function $u(\theta)$ (\ref{act})
and (\ref{der2}) we need the following matrix representation of
$\theta$ and $\partial$
  \be
  \lb{threpr}
  \begin{array}{c}
    \theta_{km} =\left(
  \begin{array}{cccc}
   0 & . & . & 0 \\
   1 & 0 & . & . \\
   . & . & 0 & . \\
   0 & . & 1 & 0
  \end{array}\right) = < k | \theta | m >, \\ \\
    \partial_{km} =\left(
  \begin{array}{ccccc}
   0 & 1 & 0 & . & 0 \\
   . & 0 & (2)_{q} & 0 & . \\
   . & . & . & . & . \\
   . & . & . & 0 & (p)_{q} \\
   0 & . & . & . & 0
  \end{array}
  \right) = < k | \partial | m >. \\
  \end{array}
  \ee
 The corresponding matrix representations of $\omega$ (\ref{gg})
 and projectors (\ref{proj}) are
  \be
  \lb{ggm}
  \begin{array}{c}
    \omega_{km} =\left(
  \begin{array}{cccc}
   1 & 0 & . & 0 \\
   0 & q & . & . \\
   . & . & . & .  \\
   0 & . & 0 & q^p
  \end{array}\right) =  < k | \omega | m >, \\ \\
    (P_n)_{km} = \delta_{kn} \, \delta_{km}, \\
  \end{array}
  \ee
 and one can obtain the relations
\be \lb{T}
\omega^T = \omega \; ,  \;\;\;
\partial^T = (\frac{\omega - 1}{q-1})\, \theta \; , \;\;\;
 \theta^T = \partial \, \left( \sum_{k=1}^p \,
\frac{1}{(k)_q} \, P_k \right) \; , 
\ee
 where $T$ is a transposition of the matrices. The last equation in
 (\ref{T}) has been obtain with the help of projectors
 (\ref{proj}).

 In eqs. (\ref{mrepr}) -- (\ref{ggm})
we have introduced the ladder of $p+1$ states $| m >$ $(m = 0,
\dots , p)$ defined by: $\partial |0 > = 0$, $\theta | m > = | m +
1 >$, $\partial | m > = (m)_q \, | m -1 >$. The dual states $< k
|$ satisfy the orthogonality condition $<k | m > = \delta_{km}$.
The matrix  $\theta_{km}$ has the form $\delta_{k,m+1}$ and in
general $(\theta^{n})_{km}= \delta_{k,m+n}$, while for the matrix
$\partial$ we have $(\partial^n)_{km} \sim \delta_{k+n,m}$. The
form of the matrices $(\theta^n)$, $(\partial^n)$ and $\omega$
corresponds to our choice of $Z_{p+1}$ grading (\ref{grad}) which
is naturally called in \cite{FIK} as "along diagonal" grading. We
use the matrix representations (\ref{threpr}), (\ref{ggm}) in our
calculations below.

 Note that the algebra (\ref{qdef1}),
 with additional relations (\ref{qdef}), represents a special
 case of the more general algebra
  \be
  \lb{cdef}
 {\cal D} \, \Theta - q \, \Theta \, {\cal D} = I \; ,
  \ee
  \be
  \lb{cdef1}
  \Theta^{p+1} = c_1 \; , \;\;\;
 {\cal D}^{p+1}= c_2
  \;\; \Leftrightarrow \;\; q^{p+1}=1 \; ,
 \ee
 where operators $c_i$ are central elements,
$deg(\Theta) = 1$ and $deg({\cal D}) = -1$. For all fixed values
of $c_i$ these algebras are isomorphic to the algebra of matrices
$Mat(p+1)$ and, therefore, the generators $\Theta, {\cal D}$ can
be expressed in terms of the elements $\theta, \, \partial$ of the
generalized Grassmann algebra (\ref{qdef1}), (\ref{qdef}). In the
case $c_i =0$ we have $\Theta = \theta$, ${\cal D} = \partial$.
Let $c_1 \neq 0$ (the case $c_2 \neq 0$ is considered below) and
operators $\Theta$ and ${\cal D}$ are
  \be
  \lb{TD}
 \Theta = \theta + \frac{1}{(p)_q !} \, \partial^p \, c_1 \; , \;\;\;
 {\cal D} =  \Theta^{-1} \, \frac{z \, \omega - 1}{q-1} \; ,
  \ee
 where $z$ is a function of central elements $c_i$
which is fixed below, $(n)_q ! := (1)_q \cdot (2)_q \cdots (n)_q$.
To find the expression for ${\cal D}$ we have used (\ref{cdef})
and the identity
 $$
{\cal D} \, \Theta - \Theta \, {\cal D} = z \, \omega   \; .
 $$
Note that the operator $\Theta$ in the matrix representation has
the familiar form of the cyclic matrix:
  $$
 \Theta_{km} =\left(
  \begin{array}{cccc}
   0 & . & 0 & c_1 \\
   1 & 0 & . & 0 \\
   . & . & 0 & . \\
   0 & . & 1 & 0
  \end{array}\right).
  $$
  Operators $\Theta$, ${\cal D}$ (\ref{TD}) automatically
 satisfy conditions (\ref{cdef}),
(\ref{cdef1}) except eq. ${\cal D}^{p+1}= c_2$ which connects the
parameter $z$ with the central elements $c_i$:
  \be
  \lb{equ1}
  c_1 \, c_2 \, (1 - q)^{p+1} = \prod_{k=0}^p \, (1- z \, q^{-k} \, \omega )=
 1 - z^{p+1},
  \ee
 where we have used that $\omega^{p+1} = 1 = q^{p+1}$.

 The case $c_2 \neq 0$ can be considered analogously
 and one can take
  \be
  \lb{TD1}
 {\cal D} = \partial + \frac{1}{(p)_q !} \, \theta^p \, c_2 \; , \;\;\;
 \Theta = \frac{z \, \omega - 1}{q-1} \, {\cal D}^{-1},
  \ee
 instead of (\ref{TD}) with the same relation
 (\ref{equ1}) on the elements $c_i$ and $z$.
 This relation is rewritten in the form of Fermat curve
 $y^{p+1} + z^{p+1} = 1$ where $y = (c_1 \, c_2)^{1/(p+1)} \, (1-q)$.
At the end of this Section we note that the differential calculus
on the generalized Grassmann algebras \cite{FIK}, \cite{FIK1} is
closely related to the differential calculus on the finite groups
$Z_n$ \cite{DMC}.

 \section{THE HEISENBERG-WEYL AND CLIFFORD ALGEBRAS}

The N-dimensional Heisenberg-Weyl algebra is generated by $2N$
operators $\theta_{i}$ and $\partial_{i}$ ($i=1,2,..,N$)
satisfying the following commutation relation

\be\lb{HW}\begin{array}{c}
  \theta_{\mu}\theta_{\nu}+\theta_{\nu}\theta_{\mu}=0, \\
  \partial_{\mu}\partial_{\nu}+\partial_{\nu}\partial_{\mu}=0, \\
  \partial_{\mu}\theta_{\nu}+\theta_{\nu}\partial_{\mu}=\delta_{\mu\nu}.
\end{array}
\ee

It is straightforward to check that \be \lb{cliff} \gamma_{\mu} =
\partial_{\mu}+\theta_{\mu} \; , \;\;\;
\gamma_{N+\mu} = i(\partial_{\mu}-\theta_{\mu}), \; \ee

(where $\mu=1,2,..,N$) is a realization of $2N$-dimensional
Euclidean Clifford algebra $[\gamma_A, \, \gamma_B ]_+ = 2
\delta_{AB}$. For our purposes it is convenient to recall the Weyl
 construction \cite{We} of the generators $\theta_\mu$,
 $\partial_\mu$ of the algebra (\ref{HW})
 \be\lb{defo}
 \theta_{\mu}=
 \underbrace{ \hat{\omega} \otimes \hat{\omega}
 \otimes....\otimes \hat{\omega}}_{(\mu-1)times}
 \otimes \hat{\theta} \otimes \underbrace{I\otimes....\otimes I}_{(N-\mu)times},
 \ee
  \be
  \lb{defo1}
 \partial_{\mu}= \underbrace{ \hat{\omega}\otimes \hat{\omega}\otimes....\otimes \hat{\omega}}_{(\mu-1)times}
 \otimes \hat{\partial}
 \otimes \underbrace{I\otimes......\otimes I}_{(N-\mu)times},
   \ee
 via the generators $\hat{\theta}$ and $\hat{\partial}$
  of the 1-dimensional fermionic Heisenberg - Weyl algebra
 (\ref{qdef1}), (\ref{qdef}) (for $p=1$)
  \be
 \lb{hw1}
 \hat{\theta} \, \hat{\partial} +
 \hat{\partial} \, \hat{\theta} = 1 \; , \;\;\;
 \hat{\theta}^2 = 0 = \hat{\partial}^2 \; , \;\;\;
 \hat{\omega} = [ \hat{\partial}, \, \hat{\theta}].
  \ee
 Using relations (\ref{defo}) and (\ref{defo1}) one can
construct the irreducible representation of the HW algebra
(\ref{HW}) and even dimensional Clifford algebras  using the
representation (\ref{threpr}), (\ref{ggm}) of the generalized
Grassmann algebra for $p=1$:
  \be
  \lb{wmatr}
 \hat{\theta} =\left(
   \begin{array}{cc}
   0 & 0 \\
   1 & 0
  \end{array}
 \right) \; , \;\;\;
    \hat{\partial} =\left(\begin{array}{cc}
   0 & 1 \\
   0 & 0
  \end{array}
 \right) \; , \;\;\;
    \hat{\omega} = \left(\begin{array}{cc}
   1 & 0 \\
   0 & -1
  \end{array}
 \right). 
  \ee

Recall that the tensorial product (in particular the tensorial
products in (\ref{defo}), (\ref{defo1})) of two matrices A and B
($n \otimes n$ and $m \otimes m$ respectively) is defined by the
following $n \cdot m \otimes n \cdot m$ matrix
  \be
  \lb{tenprod}
 A \otimes B=\left( \begin{array}{cccc}
   a_{11}B & . & . & a_{1n}B \\
   . & . & . & . \\
   . & . & . & . \\
   a_{n1}B & . & . & a_{nn}B
  \end{array}
  \right).
 \ee

 The matrices (\ref{defo}), (\ref{defo1}) are
$2^N \times 2^N$ dimensional matrices. Our task is to embed the
$2N$ dimensional Heisenberg-Weyl algebra (\ref{HW}) into the
generalized Grassmann algebra (\ref{qdef1}), (\ref{qdef}) for some
fixed $p$. It is clear (by considering the dimensions of the
matrix representations (\ref{threpr}), (\ref{ggm}) and
(\ref{defo}), (\ref{defo1})) that the Heisenberg-Weyl algebra can
be realized in terms of the q-operators only if $p+1=2^N$ and,
thus, $p$ must be odd.

  \vspace{0.5cm}
  \noindent
  {\bf Remark 1.} Matrix representations of
 the generators $\theta_\mu$ and $\partial_\mu$ (\ref{defo}),
 (\ref{defo1})), (\ref{wmatr}) satisfy conditions
  \be
  \lb{TTT}
 \theta_\mu^{T} = \partial_\mu \; , \;\;\; \partial_\mu^{T} =
 \theta_\mu \; , \;\;\;
 \theta_\mu^{*} = \theta_\mu \; , \;\;\; \partial_\mu^{*} = \partial_\mu
  \ee
where $T$ is a transposition and $*$ is a complex conjugation of
the matrices. It means that eqs. (\ref{cliff}) define the real
representation of the Clifford algebras.

  \vspace{0.5cm}
  \noindent
  {\bf Remark 2.} The Weyl construction
 (\ref{defo}), (\ref{defo1}) can be generalized for the case when
 the algebra $\{ \hat{\theta}, \hat{\partial} \}$ is taken to be
 $Z_{p+1}$ algebra (\ref{qdef1}), (\ref{qdef}) for arbitrary $p$.
 In this way one can construct multidimensional generalized
 Grassmann algebras (see \cite{FIK1}, \cite{Is} for details) and
 the algebras of covariant quantum oscillators \cite{FIK}.

 \section{$N=2,3,4$ HEISENBERG-WEYL ALGEBRAS VIA $Z_{4,8,16}$
 GENERALIZED GRASSMANN ALGEBRAS}

 \subsection{$N=2$ Heisenberg-Weyl algebra}

 To understand what we mean with embedding of algebras, let us
consider the case $N=2$ (or $p=3$). We will consider only
primitive roots $q$. It means that for $q^4=1$, we have to put
$q^2=-1$. The Weyl representation (\ref{defo}), (\ref{defo1})
gives
 \be
 \lb{We2a}
   \theta_{1}=
 \left(\begin{array}{cccc}
   0 & 0 & 0 & 0 \\
   0 & 0 & 0 & 0 \\
   1 & 0 & 0 & 0 \\
   0 & 1 & 0 & 0
 \end{array}\right) \; , \;\;\;
   \theta_{2}=\left(\begin{array}{cccc}
   0 & 0 & 0 & 0 \\
   1 & 0 & 0 & 0 \\
   0 & 0 & 0 & 0 \\
   0 & 0 & -1 & 0
 \end{array}\right), \
 \ee
 \be
\lb{We2}
  \partial_{1}=\left(\begin{array}{cccc}
   0 & 0 & 1 & 0 \\
   0 & 0 & 0 & 1 \\
   0 & 0 & 0 & 0 \\
   0 & 0 & 0 & 0
 \end{array}\right) \; , \;\;\;
  \partial_{2}=\left(
  \begin{array}{cccc}
   0 & 1 & 0 & 0 \\
   0 & 0 & 0 & 0 \\
   0 & 0 & 0 & -1 \\
   0 & 0 & 0 & 0
\end{array}\right).
\ee

 Note that the non zero elements of
$\theta_{1}$ and $\partial_1$ are in the sites $(k,k-2)$ and
$(k-2,k)$, while the non zero elements of $\theta_{2}$ and
$\partial_2$ are in the sites $(k,k-1)$ and $(k-1,k)$. It means
that the generators $\theta_{i}$ and $\partial_i$ can be expressed
(according with the along diagonal grading (\ref{grad})) in terms
of the generators of the generalized Grassmann algebra as
 \be
\lb{clin}
\theta_{1}= t_{0}\theta^2 + t_{1}\theta^3\partial \; , \;\;\;
\theta_{2}= t'_{0}\theta + t'_{1}\theta^2\partial +
t'_{2}\theta^3\partial^2 \; .
 \ee 
 \be 
 \lb{clin2}
\partial_{1}= d_{0}\partial^2 + d_{1}\theta\partial^3 \; , \;\;\;
\partial_2= d'_0\partial + d'_1 \theta \partial^2 +
d'_2 \theta^2\partial^3 \; .
  \ee
 Our aim is to determine the coefficients $t_i$, $t'_i$ and $d_i$,
 $d'_i$ in (\ref{clin}) and (\ref{clin2}) in such a way that
 generators $\theta_\mu$, $\partial_\nu$ obey the defining
 relations (\ref{HW}). According with the representation
 (\ref{ggm}) we find that
  \be
  \lb{clin3}
 \theta_1 = \theta^2 \; , \;\;\; \theta_2 = \theta \, (P_0 - P_2) =
 (P_1 - P_3) \, \theta \; .
  \ee
 Since the generators $\partial_\mu$ are related to the generators
 $\theta_\mu$ by the transpositions (\ref{TTT}) and using (\ref{T})
 one can deduce from (\ref{clin3}) the expressions
  \be
  \lb{clin4}
    \partial_1 =  \frac{1}{(2)_q} \, \partial^2 \, \left( P_2  +
 \frac{1}{(3)_q} \, P_3 \right) \; , \;\;\;
    \partial_2 =  \partial \, \left( P_2  -
 \frac{1}{(3)_q} \, P_3 \right) \; . 
  \ee

 Since $q$ is a primitive root of the equation $q^4 =1$ the
 following identities hold:
  \be
  \lb{clin5}\begin{array}{c}
    q^2 = -1 \; , \;\;\; (2)_{q}= 1 + q \; , \;\;\; (3)_{q}= 1 + q +
 q^2 = q  \\
    \Downarrow \\
    (2)_{q} \, (3)_{q} = (1 + q) \, q = q-1, \\
  \end{array}
  \ee
 and using relations (\ref{qdef1}), (\ref{usr}) one can deduce
 another representation for the generators of the $N=2$
 Heisenberg-Weyl algebra
 $$
 \theta_{1} = \theta^2 \; , \;\;\; \theta_{2}= \theta -\theta^2 \,
 \partial + q \, \theta^3 \, \partial^2,
 $$
 $$
 \partial_{1}= \frac{1}{2} \, (1-q) \, \partial^2 - \theta\partial^3 \; ,
\;\;\;
 \partial_2= \partial -\theta \partial^2 + (1 + q) \, \theta^2
 \partial^3.
 $$
 One can express these generators in terms of the operator $\omega$
 using (\ref{clin3}), (\ref{clin4}) and (\ref{nilp})
 $$
 \theta_{1} = \theta^2 \; , \;\;\;
   \partial_{1}= (\frac{\omega - 1 -q}{1-q})\, \partial^2 \; , \;\;\;
   \theta_{2}= \theta \, (\frac{\omega - q}{1-q})^2 \; , \;\;\;
   \partial_2 = \omega \, (\frac{\omega - q}{1-q}) \, \partial.
 $$

 \subsection{$N=3$ Heisenberg-Weyl algebra}

 Using Weyl construction (\ref{defo}), (\ref{defo1}) one can obtain
 the following matrix representations for the generators
 $\theta_\mu$ $(\mu = 1,2,3)$:
  \be
  \lb{cli61}\begin{array}{c}
   \theta_1 = \sum_{m=0}^{3} \, e_{m+4,m}, \\
   \theta_2 = \sum_{m=0}^{1} \, ( e_{m+3,m} - e_{m+6,m+3} ), \\
   \theta_3 = e_{1,0} - e_{3,2} - e_{5,4} + e_{7,6}, \\
  \end{array}
  \ee
 where the matrix units $e_{ij}$ have been introduced
  \be
  \lb{unit}
  (e_{ij})_{km} = \delta_{ik} \, \delta_{jm} \; , \;\;\;
  e_{ij} \, e_{kl} = \delta_{jk} \, e_{il}. \; .
  \ee
 According with the representation (\ref{ggm}) we find that
  \be
  \lb{clin62}
 \theta_1 = \theta^4  \; , \;\;\;
 \theta_2 = \theta^2 \, (P_0 + P_1 -P_4 - P_5)= (P_2 + P_3 - P_6 - P_7) \, \theta^2 \; , \;\;\;
 \theta_3 = \theta \, (P_0 - P_2 - P_4 + P_6) = (P_1 - P_3 - P_5 + P_7) \, \theta.
  \ee
 Applying the transpositions (\ref{TTT}) and using (\ref{T}) one
 can deduce from (\ref{clin62}) the expressions for generators
$\partial_\mu$:
  \be
  \lb{clin63}
  \begin{array}{l}
 \partial_1 = \partial^4( \sum_{k=4}^7
 \frac{1}{(k-3)_q(k-2)_q(k-1)_q(k)_q}) P_k , \\
 \partial_2 = \partial^2 (\sum_{k=2,3}
 - \sum_{k=6,7})\frac{(k-2)_q!}{(k)_q!}P_k , \\
 \partial_3 = \partial( \sum_{k=1,7}
 - \sum_{k=3,5}) \frac{1}{(k)_q}P_k.
  \end{array}
  \ee

 \subsection{$N=4$ Heisenberg-Weyl algebra}

 Using Weyl construction (\ref{defo}), (\ref{defo1}) one can obtain
 the following matrix representations for the generators
 $\theta_\mu$ $(\mu = 1,2,3)$:
  \be
  \lb{cli81}
    \begin{array}{l}
  \theta_1 = \sum_{m=0}^{7} \, e_{m+8,m}, \\

  \theta_2 = (\sum_{m=0}^{3} -\sum_{m=8}^{11}) \,  e_{m+4,m},
  \\
  \theta_3 = (\sum_{m=0,1,12,13} -\sum_{m=4,5,8,9}) \,  e_{m+2,m},
  \\
 \theta_4 = (\sum_{m=0,6,10,12} -\sum_{m=2,4,8,14}) \,  e_{m+1,m}.
  \end{array}
  \ee
 According with the representation (\ref{ggm}) we find that
  \be
  \lb{clin82}
   \begin{array}{l}
 \theta_1 = \theta^8 \; ,  \\
  \theta_2 =  (\sum_{m=4}^{7} -\sum_{m=12}^{15}) \, P_{m} \theta^4 \; , \\
 \theta_3 =  (\sum_{m=2,3,14,15} -\sum_{m=6,7,10,11})  \, P_{m}
 \theta^2 \; , \\
 \theta_4 =  (\sum_{m=1,7,11,13} -\sum_{m=3,5,9,15})  \, P_{m}
 \theta.
  \end{array}
  \ee
 Applying the transpositions (\ref{TTT}) and using (\ref{T}) one
 can deduce from (\ref{clin62}) the expressions for generators
 $\partial_\mu$:
  $$
  \begin{array}{l}
 \partial_1 =   \partial^8 \, ( \sum_{k=8}^{15} \,
 \frac{(k-6)_q!}{(k)_q!} \, P_k ) , \\
 \partial_2 =  \partial^4 \, ( \sum_{k=4}^7  -
\sum_{k=12}^{15} ) \, \frac{(k-4)_q!}{(k)_q!} \, P_k  , \\
 \partial_3 =  \partial^2 \, ( \sum_{k=2,3,14,15}  -
\sum_{k=6,7,10,11}) \, \frac{(k-2)_q!}{(k)_q!} \,P_k  , \\
 \partial_4 =  \partial \,( \sum_{k=1,7,11,13} -
 \sum_{k=3,5,9,15}) \, \frac{1}{(k)_q} \, P_k  .\\
 \end{array}
 $$

\section{ EXAMPLE: $N=2$ SKDV EQUATIONS IN TERMS OF
  $Z_4$ GENERALIZED GRASSMANN ALGEBRA}

  We start with the discussion of the $N=1$ super-extension of the KdV equation
 \be
 \frac{\partial \Phi}{\partial t} = -\partial_x \Big (( \partial^{2}_x \Phi)
+3\Phi ({\cal D}\Phi)\Big ).
 \ee
  It is known (see e.g.\cite{MR}) that the Lax operator $L$ with the Lax pair
representation
  in this case can be taken in the form
  \be
  \lb{1kdv}
  L = \partial_x^2 + \Phi \, {\cal D} \ \quad;
  \quad  \frac{\partial L}{\partial t} =\Big [L,L^{\frac{3}{2}}_{\geq 0} \Big ]
,
  \ee
 where $\geq 0$
denotes the projection   onto the purely (super)differential part
of the operator, $\partial_x \equiv
 \frac{\partial}{\partial x}$ and we have used the odd superfield
 $\Phi(x,\theta) = \psi(x) + \hat{\theta} \, u(x)$
 and covariant super-derivative
 ${\cal D} = \hat{\partial} + \hat{\theta} \, \partial_x$.

  The  operators $\hat{\theta}$, $\hat{\partial}$ are the same as in
(\ref{hw1}),
  (\ref{wmatr}). Since the fermionic field $\psi(x)$ anticommutes
  with $\hat{\theta}$ and $\hat{\partial}$ (\ref{wmatr}) we should represent
  it in the form
  \be
  \lb{2kdv}
  \psi(x) = \hat{\omega} \, \hat{\psi}(x) =
   \left(\begin{array}{cc}
   \hat{\psi} & 0 \\
   0 & -\hat{\psi}
  \end{array}
 \right) \; ,
  \ee
  which guaranties the anticommutation of $\psi$ with $\hat{\theta}$ and
  $\hat{\partial}$. Substitution of the matrix representations
  (\ref{wmatr}) and (\ref{2kdv}) in (\ref{1kdv}) gives the
  matrix representation for the $N=1$ SKdV Lax operator
  \be
  \lb{3kdv}
  L = \partial_x^2 +
   \left(\begin{array}{cc}
   \hat{\psi} & 0 \\
   u & -\hat{\psi}
  \end{array} \right) \,
   \left(\begin{array}{cc}
   0 & 1 \\
   \partial_x & 0
  \end{array}
  \right)
     = \partial_x^2 +
   \left(\begin{array}{cc}
   0 & \hat{\psi}  \\
    -\hat{\psi}\partial_x & u
  \end{array} \right). 
 \ee

 Here the field $\hat{\psi}(x)$ should be considered as a fermionic
field while $u(x)$ is a bosonic Virasoro field. Thus, we
represent the Lax operator $L$ for super-KdV equation in two
different but absolutely equivalent forms. The first formula
(\ref{1kdv}) gives the representation of the $L$-operator in terms
of $Z_2$ graded algebra (\ref{qdef1}), (\ref{qdef}), while the
second formula (\ref{3kdv}) gives corresponding graded matrix
realization.

Now the discussion of the $N=2$ SKdV Lax operators is in order. It
is well known that there are three different completely integrable
$N=2$ supersymmetric extensions of the Korteweg-de Vries equations
\cite{ZP} \cite{ZP1}. All these extensions could be written as
  \be
 \lb{lp}
 \frac{\partial \Phi}{\partial t} =
 \partial_x \, \Big ( -(\partial^{2}_x\Phi) + 3\Phi({\cal D}_1{\cal D}_2\Phi)
+ \frac{1}{2}(\alpha -1)({\cal D}_1{\cal D}_2\Phi^2) +
\alpha\Phi^{3} \Big ),
\ee
 where $\alpha=4,-2,1$ and $\Phi(x, \theta_1,\theta_2) = w(x) + \theta_1 \,
\psi_1(x) +
 \theta_2 \, \psi_2(x) + \theta_2 \, \theta_1 \, u(x) $.
 We would like to show that the Lax operators for these generalizations could
be written in terms
 of the $Z_4$ generalized Grassmann algebra. Let us notice that  the Lax
operators for
 the $\alpha=1$ case is
   \be
   \lb{kdv1}
 L_{1} = \partial_x - \partial^{-1}_x \, {\cal D}_1 \, {\cal D}_2 \,
 \cdot \Phi(x, \theta_1,\theta_2) \; ,
   \ee
  where the superderivatives ${\cal D}_i$  are
  \be
  \lb{kdv2}
  \begin{array}{c}
  {\cal D}_1 = \partial_1 + \theta_1 \, \partial_x \; , \;\;\;
  {\cal D}_2 = \partial_2 + \theta_2 \, \partial_x .
  \end{array}
  \ee
 The supersymmetric KdV equation is obtained from the Lax pair
 representation as
  \be
  \lb{lp1}
  \frac{\partial L_1}{\partial t} =
 \left [ L_1,(L^{3}_{1})_{\geq 0} \right] \ee
  The Lax operators for the $\alpha=4,-2$ are
 described in terms  of the previous operator $L_1$ as
  \be
 L_{-2}=L^{\dagger}_{1} \; L_{1} \quad , \quad L_4=( L^{\dagger}_1 \; L_{1}
)_{\geq 0}
 \; ,
  \ee
 where $\dagger$ denotes the hermitian conjugation. From that
 reason we consider therefore the case $\alpha=1$ only.

 The component $u(x)$ is a Virasoro field and variables
 $\theta_i$ and $\psi_j$ anticommute with each other. It means that
 we can represent $\psi_i$ in the form
  $$
 \psi_i = (\hat{\omega} \otimes \hat{\omega} ) \cdot \hat{\psi}_i =
  \left(\begin{array}{cccc}
   \hat{\psi}_i & 0 & 0 & 0 \\
   0 & -\hat{\psi}_i & 0 & 0 \\
   0 & 0 & -\hat{\psi}_i & 0 \\
   0 & 0 & 0 & \hat{\psi}_i
 \end{array}\right)
 $$
 $$
 [\theta_i , \, \hat{\psi}_j]= 0 =
 \{ \hat{\psi}_i , \, \hat{\psi}_j \}_+ .
 $$
 In matrix notations we have
  \be
  \lb{DD}
  \begin{array}{c}
    {\cal D}_1 = \left(\begin{array}{cccc}
   0 & 0 & 1 & 0 \\
   0 & 0 & 0 & 1 \\
   \partial_x & 0 & 0 & 0 \\
   0 & \partial_x & 0 & 0
  \end{array}\right) \; , \;\;\;
    {\cal D}_2 = \left(\begin{array}{cccc}
   0 & 1 & 0 & 0 \\
   \partial_x & 0 & 0 & 0 \\
   0 & 0 & 0 & -1 \\
   0 & 0 & -\partial_x & 0
  \end{array}\right), \\ \\
  {\cal D}_1 \, {\cal D}_2 =
  \left(\begin{array}{cccc}
   0 & 0 & 0 & -1 \\
   0 & 0 & -\partial_x & 0 \\
   0 & \partial_x & 0 & 0 \\
   \partial_x^2 & 0 & 0 & 0
 \end{array}\right) \; , \;\;\;
   \Phi = \left(
 \begin{array}{cccc}
   w & 0 & 0 & 0 \\
   \hat{\psi}_2 & w & 0 & 0 \\
    \hat{\psi}_1 & 0 & w & 0 \\
   - u & - \hat{\psi}_1 & \hat{\psi}_2 & w
  \end{array}\right). \
 \end{array}
 \ee

 Finally we write the Lax operator (\ref{kdv1}) in the matrix form
  \be
  \lb{laxm}
   L =  \partial_x \, {\bf 1}
  - \left(\begin{array}{cccc}
   0 & 0 & 0 & -\partial_x^{-1} \\
   0 & 0 & -1 & 0 \\
   0 & 1 & 0 & 0 \\
   \partial_x & 0 & 0 & 0
 \end{array}\right)
 \times 
   \left(\begin{array}{cccc}
   w & 0 & 0 & 0 \\
   \hat{\psi}_2 & w & 0 & 0 \\
    \hat{\psi}_1 & 0 & w & 0 \\
   - u & - \hat{\psi}_1 & \hat{\psi}_2 & w
  \end{array}\right).
  \ee

 The superfield $\Phi$ and covariant derivatives ${\cal D}_i$ in
 terms of $Z_4$ variables are represented in the form
  \be
  \lb{kdv6}
    \Phi = w(x) +  \theta \, (P_0 + P_2) \, \hat{\psi}_2(x) 
    + \theta^2 \, (P_0 - P_1) \, \hat{\psi}_1(x) - \theta^3 \, u(x). 
  \ee
  \be
  \lb{kdv4}
  \begin{array}{c}
   {\cal D}_1 = \frac{1}{(2)_q} \, \partial_\theta^2 \left( P_2 +
 \frac{1}{(3)_q} \, P_3 \right) + \theta^2 \, \partial_x \\
    {\cal D}_2 =
  \partial_\theta \left( P_1 - \frac{1}{(3)_q} \, P_3 \right) +
   \theta \, (P_0 - P_2) \, \partial_x,  \\
  \end{array}
  \ee
 and using these relations we obtain
  \be
  \lb{kdv5}
  \partial_x^{-1} \, {\cal D}_1  \, {\cal D}_2 = \frac{(q+1)}{2} \,
  \partial_\theta^3 \, \partial_x^{-1} +
  \frac{(q-1)}{2} \, \partial_\theta  \, P_2
  + \theta  \, P_1  + \theta^3 \, \partial_x .
  \ee

  Let us introduce the $Z_4$ covariant derivative
  \be
  \lb{kdv3}
    {\cal D} = \partial_\theta + \frac{1}{(3)_q!} \, \theta^3 \,
 \partial_x =
 \partial_\theta - \frac{1}{2} \, (q+1) \, \theta^3 \, \partial_x \; , \;\;\;
    {\cal D}^4 = \partial_x \; , 
  \ee
 where the operators $\theta$ and $\partial_\theta$ generate the
 $Z_4$ algebra (\ref{qdef1}), (\ref{qdef}) for $p=3$ and parameter
 $q$ satisfy $q^2 = -1$. Now we express all $\partial_\theta^k$ via
 operator $\theta$ and $Z_4$- covariant derivative ${\cal D}$:
  $$
  \begin{array}{l}
  \partial_\theta = {\cal D} + \frac{1}{2} \, (q+1) \, \theta^3 \,
  \partial_x \; ,  \\ \\
  \partial_\theta^2 = {\cal D}^2 + \left( q \, {\cal D} \, \theta^3  +
  \frac{(q+1)}{2} \, \theta^2 \right) \,  \partial_x \; , \\ \\
  \partial_\theta^3 = {\cal D}^3 + \left( \frac{(q-1)}{2} \, {\cal D}^2 \,
\theta^3 +
  \frac{(q-1)}{2} \, {\cal D} \, \theta^2   + q \, \theta \right) \,
\partial_x \; .
  \end{array}
  $$
  Using this relation the differential operator (\ref{kdv5}) is rewritten as
  $$
 \left\{ \frac{q+1}{2} \, \partial_x^{-1}  \,
  {\cal D}^3 - \frac{1}{2} \, {\cal D}^2 \, \theta^3 +
  {\cal D}(\frac{(q-1)}{2} \, P_2
  - \frac{1}{2} \, \theta^2)
  + \theta \left( P_1 + \frac{(q-1)}{2} \right)
   + \theta^3 \, \partial_x \right\} \; .
  $$

 {}Finally we obtain the expression for the Lax operator $L$
 (\ref{laxm}) as a pseudodifferential operator with respect to the
 fractional covariant derivative ${\cal D}$:
  \be
  \lb{kdv7}
  L = \partial_x + {\cal D}^{-1} \cdot u_{-1} + u_0 +
  {\cal D} \cdot u_{1} + {\cal D}^{2} \cdot u_{2} + \partial_x \cdot u_{4}
  \; ,
   \ee
  where the fractional superfields $u_k(x, \theta)$
  are defined in terms of the unique fractional prepotential $\Phi$
(\ref{kdv6}).
  $$
  u_{-1} = \frac{1}{q-1} \, \Phi \; , \;\;\;  u_{0} = \theta \,
  \left( \frac{1}{q+1} - P_1 \right) \, \Phi \; ,
  $$
  $$
  u_{1} = \left( \frac{1-q}{2} \, P_2 + \frac{1}{2} \, \theta^2 \right) \, \Phi
\; ,
  \;\;\;   2 \, u_2 = - u_{4} = \theta^3  \, \Phi \; .
  $$

 \section{CONCLUSION}

 In this parer we have shown that the fermionic Heisenberg-Weyl algebra with
$2N=D$ fermionic generators is equivalent to the generalized
Grassmann algebra with two fractional generators of the order of
the parastatistic: $p = 2^{N} -1$. The 2,3 and 4 dimensional
Heisenberg - Weyl algebras (they are related to the
4,6 and 8 dimensional Clifford algebras) were  explicitly given. The $N=2,3,4$
extended supersymmetries can be described in terms of these
algebras. We reformulate the Lax approach for the supersymmetric
Korteweg - de Vries equation in terms of the generators of the
generalized Grassmann algebra. Note that the theory of the
pseudodifferential operators on the $(1|1)$ superline \cite{MR}
can be directly generalized to the theory of pseudodifferential
operators on the fractional superline $(x, \theta)$ (we plan to
return to these problems in one of the forthcoming papers). On the
other hand, in the similar manner,  one can  construct the
$Z_{p+1}$- graded extensions of the KdV and Kadomtsev-Petviashvily
hierarchies. These extensions should be probably  related to the
graded matrix generalizations of the KdV and KP equations.

\section*{Acknowledgments}
We would like to thank Dr. A. Sorin for the fruitful discussions.
The work of API was partially supported by the RFBR grant No.
00-01-00299 and Bogoliubov - Infeld Program.


\begin{thebibliography}{99}

 \bibitem{ABL} C.Ahn, D.Bernard and A.LeClair, \it Nucl.Phys. \bf B346 \rm (1990) 409.

 \bibitem{FIK} A.T. Filippov, A.P. Isaev and A.B. Kurdikov, \it Mod. Phys. Lett. {\bf A7} \rm (1992) 2129.

 \bibitem{Dur0} S. Durand, \it Mod. Phys. Lett. {\bf A7} \rm (1992) 2905.

 \bibitem{FIK2} A.T.Filippov, A.P.Isaev and A.B.Kurdikov,
 \it Int. J. Mod. Phys. {\bf A8} \rm (1993) 4973,
 {\tt hep-th/9212157}.

 \bibitem{AzM} J.A. de Azcarraga and A.J. Macfarlane, \it J. Math. Phys. {\bf 37} \rm (1996) 1115.

 \bibitem{AD} H. Ahmedov and O.F. Dayi, \it Mod. Phys. Lett. {\bf A15} 
\rm (2000) 1801, {\tt math.qa/9905164}.

 \bibitem{Dur} S.Durand, \it Phys. Lett. {\bf B 312} \rm (1993) 115; R.S. Dunne, A.J. Macfarlane, J.A. de Azcarraga,
 J.C. Perez Bueno, \it Int. J. Mod. Phys. {\bf A12} \rm (1997) 3275, {\tt hep-th/9610087}; M.Daoud and M. Kibler,
 "A fractional supersymmetric oscillator and its coherent states", {\tt math-ph/9912024}; F. Kheirandish and M.
Khorrami, \it Eur. Phys. J. {\bf C20} \rm (2001) 593-597, {\tt
hep-th/0007073}; N.Fleury, M. Rausch de Traubenberg, \it Mod. Phys.
Lett. {\bf A11} \rm (1996) 899,{\tt hep-th/9510108};
M.~Rausch de Traubenberg and P.~Simon,
Nucl.\ Phys.\ B {\bf 517} (1998) 485; {\tt hep-th/9606188};
A.~Mostafazadeh, J.\ Phys.\ A {\bf 34} (2001) 8601,
{\tt math-ph/0110013}.

\bibitem{Isx} A.P. Isaev, ``Cyclic Paragrassmann Representations for Covariant Quantum Algebras'',
in Proc. of 2nd M. Born Symposium, ``Spinors, Twistors and Clifford Algebras'', A. Borowiec a.o. (Eds.),
Kluwer Acad. Publishers (1993), pp. 309-316.


 \bibitem{Is} A.P. Isaev, \it Int. J. Mod. Phys. {\bf A} \rm 12
 (1997) 201, {\tt q-alg/9609030}.

 \bibitem{DVT} M. Dubois-Violette and I.T. Todorov, \it Lett. Math. Phys.,
 {\bf 42} \rm (1997) 183, {\tt hep-th/9704069};
  M. Dubois-Violette et al., "A finite Dimensional Gauge Problem in
  the WZNW Model", {\tt hep-th/9910206}; 
P. Furlan et al., "Quantum Matrix algebra for the $SU(N)$ WZNW Model", {\tt hep-th/0003210}; 
P. Furlan, L. Hadjiivanov and I.T. Todorov, J. Phys. {\bf A 34} (2001) 4857.

\bibitem{Ritt}
P.F. Arndt, T. Heinzel and V. Rittenberg, J. Phys. A: Math. Gen.
{\bf 31} (1998) 833, {\tt cond-mat/9703182};
F.C. Alcaraz, S. Dasmahapatra and V. Rittenberg, J. Phys. A: Math. Gen.
{\bf 31} (1998) 845, {\tt cond-mat/9705172};
A.P. Isaev, P.N. Pyatov and V. Rittenberg, J. Phys. A: Math. Gen.
{\bf 34} (2001) 5815, {\tt cond-mat/0103603}.

 \bibitem{Dobr} M. Angelova, V.K. Dobrev and A. Frank, \it J. Phys. {\bf A34} \rm (2001) L503, {\tt cond-mat/0105193}.

 \bibitem{ArCo} M. Aric and D.D. Coon, \it J. Math. Phys. {\bf 17} \rm,
  (1976) 524.

 \bibitem{MaBi} A.J.Macfarlane, \it J. Phys. A: Math. Gen. {\bf 22}
 \rm (1989) 4581; L.C.Biedenharn, \it J. Phys. A: Math. Gen. {\bf 22} \rm (1989) L873.


  \bibitem{DMC} A. Dimakis and F. Muller - Hoissen, \it J. Phys. {\bf A 27} \rm (1994) 3159,
{\tt hep-th/9401149}; L. Castellani and Ch. Pagani, Finite Group
Discretization of Yang - Mills and Einstein Actions, Preprint
DFTT-26-01 (2001), {\tt hep-th/0109163}.

 \bibitem{We} R. Brauer and H. Weyl, \it Amer. Journ. Math.{\bf 57} 
\rm (1935) 425.

 \bibitem{FIK1} A.T. Filippov, A.P. Isaev and A.B. Kurdikov,
 \it Teor. Math. Phys.{\bf 94} \rm (1993) 213.

 \bibitem{ZP} C. Laberge, P. Mathieu, \it Phys. Lett.{\bf 215B} \rm (1988) 718;
 W. Oevel, Z. Popowicz, \it Comm. Math. Phys. {\bf 139} \rm (1991)
 441.

 \bibitem{ZP1} Z. Popowicz, \it Phys. Lett {\bf A174} \rm (1993) 411.

 \bibitem{MR} Yu.I. Manin and A.O. Radul, \it Comm. Math. Phys. {\bf
 98} \rm (1985) 65.



 \end{thebibliography}
 \end{document}